\documentclass[sigconf]{acmart}

\usepackage{booktabs} 
\usepackage{multirow}
\usepackage{stfloats}
\usepackage{xr}

\usepackage{amsmath}
\DeclareMathOperator*{\argmax}{arg\,max}
\usepackage{makecell}
\setcellgapes{1.5pt}
\usepackage{graphicx}
\usepackage{balance}

\setcopyright{rightsretained}

\copyrightyear{2018} 
\acmYear{2018} 
\setcopyright{acmlicensed}
\acmConference[MM '18]{2018 ACM Multimedia Conference}{October 22--26, 2018}{Seoul, Republic of Korea}
\acmBooktitle{2018 ACM Multimedia Conference (MM '18), October 22--26, 2018, Seoul, Republic of Korea}
\acmPrice{15.00}
\acmDOI{10.1145/3240508.3240707}
\acmISBN{978-1-4503-5665-7/18/10}

\begin{document}
\title{Deep Multimodal Image-Repurposing Detection}

\author{Ekraam Sabir}
\orcid{1234-5678-9012}
\affiliation{%
  \institution{USC Information Sciences Institute}
  \streetaddress{4676 Admiralty Way}
  \city{Marina Del Rey}
  \state{CA}
  \postcode{90292}
}
\email{esabir@isi.edu}

\author{Wael AbdAlmageed}
\affiliation{%
  \institution{USC Information Sciences Institute}
  \streetaddress{4676 Admiralty Way}
  \city{Marina Del Rey}
  \state{CA}
  \postcode{90292}
}
\email{wamageed@isi.edu}

\author{Yue Wu}
\affiliation{%
  \institution{USC Information Sciences Institute}
  \streetaddress{4676 Admiralty Way}
  \city{Marina Del Rey}
  \state{CA}
  \postcode{90292}
  }
\email{yue_wu@isi.edu}

\author{Prem Natarajan}
\affiliation{%
  \institution{USC Information Sciences Institute}
  \streetaddress{4676 Admiralty Way}
  \city{Marina Del Rey}
  \state{CA}
  \postcode{90292}
}
\email{pnataraj@isi.edu}

\renewcommand{\shortauthors}{E. Sabir et al.}

\begin{abstract}
Nefarious actors on social media and other platforms often spread rumors and falsehoods through images whose metadata (e.g., captions) have been modified to provide visual substantiation of the rumor/falsehood. This type of modification is referred to as image repurposing, in which often an unmanipulated image is published along with incorrect or manipulated metadata to serve the actor's ulterior motives. We present the Multimodal Entity Image Repurposing (MEIR) dataset, a substantially challenging dataset over that which has been previously available to support research into image repurposing detection.  The new dataset includes location, person, and organization manipulations on real-world data sourced from Flickr. We also present a novel, end-to-end, deep multimodal learning model for assessing the integrity of an image by combining information extracted from the image with related information from a knowledge base.  The proposed method is compared  against state-of-the-art techniques on existing datasets as well as MEIR, where it outperforms existing methods across the board, with AUC improvement up to 0.23.
\end{abstract}

%
%



\maketitle

\section{Introduction}
\begin{figure*}[h]
\centering
\includegraphics[width=\textwidth]{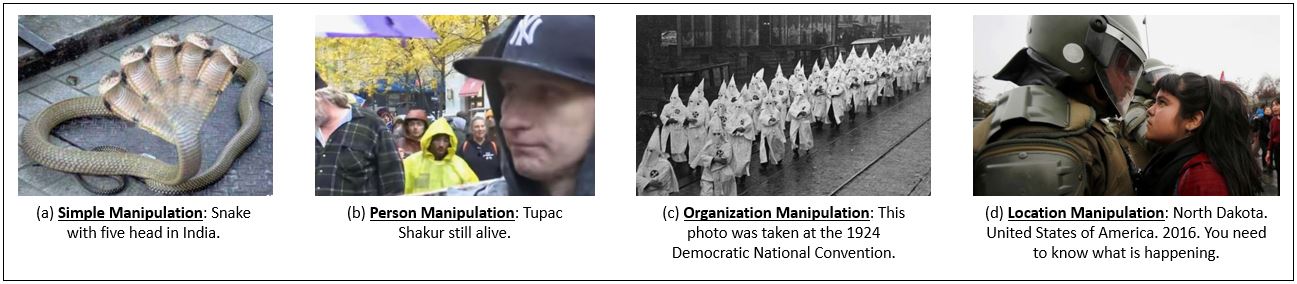}
\caption{Examples of real-world manipulated information. Image (a) is an obvious hoax with digital image manipulation and is less likely to fool people. Image (b) falsely claims that Tupac Shakur is alive. In image (c), Democratic Party is incorrectly referenced. Image (d) misattributes the photograph to a protest at the North Dakota Pipeline, when it was originally taken in Chile. Images (b), (c) and (d) are not digitally manipulated, but the details accompanying them are, making detection more challenging.}
\label{fig:internet_misinformation}
\end{figure*}

Social media is increasingly becoming the news source of choice, replacing conventional news outlets. In \cite{marchi_facebook_2012}, Marchi \textit{et al.} present a survey, showing that teens are increasingly inclined towards evolving news articles over conventional and objective renderings of journalism. However, the  unregulated use of social media makes it convenient to float rumors or present misleading information. Reasons for spreading misleading information can vary with possible malicious intent, political motivation or simply due to ignorance, as shown in Figure \ref{fig:internet_misinformation}. The left-most example is relatively easy for an educated audience to identify as fake. The right three examples, however, highlight the difficulty of detecting manipulated image data when the details being manipulated are subtle, subliminal and often interleaved with the truth. The right three images each have one entity manipulated in the associated caption, while the image remains untampered. In Figure \ref{fig:internet_misinformation}(b) the deceased rapper Tupac Shakur is incorrectly referenced. Figure \ref{fig:internet_misinformation}(c) has the Democratic National Convention referenced instead of the Ku Klux Klan, while the image in Figure \ref{fig:internet_misinformation}(d) is from a protest in Chile instead of North Dakota. It is noteworthy that the manipulated details in the right three examples are semantically coherent and leave no digital footprint in the image itself, making their detection difficult.

In the broad category of manipulation detection, the research community has invested significantly in identifying digital manipulations of images \cite{qureshi_bibliography_2015}, which involves identifying whether an image has been tampered with at the pixel and/or object level. However, cases of image repurposing---in which the image is authentic but the accompanying metadata has been manipulated---have not been thoroughly studied. The right three images in Figure \ref{fig:internet_misinformation} are examples of image repurposing.

There are crowd-sourced attempts to vet suspicious information spreading on social media. Facebook has recently launched a reporting and fact checking feature on its platform. Snopes\footnote{https://www.snopes.com/} is another fact checking website which debunks hoaxes and fake news. However the reach of these efforts is limited and requires significant human time, resource, motivation and skill. Large-scale detection of image repurposing necessitates the development of  automatic systems that could vet the credibility of an image from internet sources itself.

In this paper we attempt to address these challenges by developing a deep multimodal system for image repurposing detection. To the best of our knowledge, there has been one previous attempt at image repurposing detection in literature \cite{jaiswal_multimedia_2017}, in which the authors pose the problem as a semantic integrity assessment problem, where the semantics are potentially inconsistent between images and captions in a query \emph{package}. The package is defined as an image and associated text caption. However, semantically inconsistent manipulations are farther from real-world examples as demonstrated in Figure \ref{fig:internet_misinformation}. We use the definition of the package introduced in \cite{jaiswal_multimedia_2017}. Meanwhile, we also extend it to include the global positioning system (GPS) information included in the image metadata. We call this a \emph{multimodal image package}.

The contributions of this paper are:
\begin{enumerate}
\item A new and challenging dataset with relatively intelligent, labeled and semantically coherent manipulations over what is currently available for research.\footnote{https://github.com/Ekraam/MEIR}
\item A novel deep multimodal model that uses multitask learning for image repurposing detection
\end{enumerate}

The remainder of this paper is organized as follows: Section \ref{sec:related_works} discusses related works. Section \ref{sec:data} discusses the dataset and procedure involved in creating it. Section \ref{sec:method} describes our approach to image repurposing detection and the model, and Section \ref{sec:evaluation} gives experimentation details and describes baselines.

\section{Related Work}\label{sec:related_works}

In \cite{zampoglou_web_2016}, Zampoglou \textit{et al.} identify the importance of vetting images and accompanying information for journalistic review purposes. They rely on existing research in developing a graphical user interface (GUI) which can be used to verify the content of information found on the web. However, they are limited to discovering digital manipulations of images, which has been the focus of the research community \cite{singh_copy_2016}\cite{wu_deep_2017}\cite{qureshi_bibliography_2015}\cite{asghar_copy-move_2017}. Digital manipulations are mainly divided into copy-move and image-splicing problems. Copy-move involves duplicating objects within the same image, while splicing inserts objects from a foreign image. In \cite{asghar_copy-move_2017}, Asghar \textit{et al.} broadly classify digital image manipulation detection techniques into active and passive (blind) methods. Active methods involve a signature or watermark which has to be extracted from the image. The blind or passive approach looks for inconsistencies and artifacts in the image being investigated, which might result from digital manipulations.

To the best of our knowledge, there has been a single attempt at image repurposing detection---by Jaiswal \textit{et al.} in \cite{jaiswal_multimedia_2017}. The authors present the concept of a multimedia package, which in theory contains an untampered image and related metadata, but is limited to image caption pairs in their paper. They induce manipulations in packages by replacing captions from random packages. They also use an unlabeled knowledge base which is assumed to comprise untampered multimedia packages. The overall method involves learning joint representations from the knowledge base and performing anomaly detection on representations of query packages. The motivation for this approach is that tampering in query packages will manifest itself as an outlier compared to distribution of representations from the knowledge base.

The approach in the above formulation is useful for laying a foundation in image repurposing detection. However, it suffers from some issues. First, manipulations are semantically incorrect across images and captions, making them relatively easier to detect. Putting it in context, an image with an elephant and a manipulated caption describing a dining table are two widely different object classes and therefore relatively easy to discern. Second, as seen from examples in Figure \ref{fig:internet_misinformation}, actual image repurposing involves manipulating a detail which is semantically coherent, but misrepresents the image. Manipulations involving swapped captions in \cite{jaiswal_multimedia_2017} are farther from such real-world examples. Third, the knowledge base provided does not contain information related to the query package. It is useful in an anomaly detection framework, but is not adaptable when related information is crucial for image repurposing detection. We address some of these issues in this work.

\citet{ruchansky_csi:_2017} provide an excellent summary of rumor detection literature. The work by Jin \textit{et al.} in \cite{jin_multimodal_2017} tackles rumor detection and continues a similar body of work \cite{jin_mcg-ict_2015}\cite{jin_novel_2017}. It performs rumor detection on image caption pairs from Weibo and Twitter. The dataset is described by Boididou \textit{et al.} in \cite{boididou_verifying_2015}. However, it differs from our work in two key aspects. First, spliced and photoshopped images are included in their dataset, which have already been studied. They are qualitatively closer to  Figure \ref{fig:internet_misinformation}(a) and are easier to detect for an informed audience. Second, the attention-based model presented in their work relies solely on the information inconsistency visible in a query package to detect fake news. Similar to the setup in \cite{jaiswal_multimedia_2017}, this approach does not account for semantically coherent manipulations within query information.
\section{Multimodal Entity Image Repurposing Dataset (MEIR)}\label{sec:data}

\begin{table*}[t]
\small
\makegapedcells
\begin{center}
\begin{tabular}{|c|c|c|c|}
\cline{2-4}
\multicolumn{1}{c|}{} &
\multicolumn{2}{c|}{\textbf{Reference Dataset}} & \textbf{Manipulated Dataset} \\
\cline{2-4}
\multicolumn{1}{c|}{} & \textbf{Related Package} & \textbf{Manipulation Source Package} & \textbf{Repurposed Query Package} \\
\cline{2-4}
\hline
\multirow{1}{*}[9ex]{\rotatebox[origin=c]{90}{Image}} & 
\includegraphics[height=2cm,keepaspectratio]{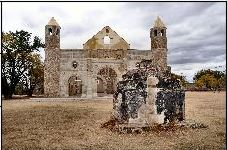} & \includegraphics[height=2cm,keepaspectratio]{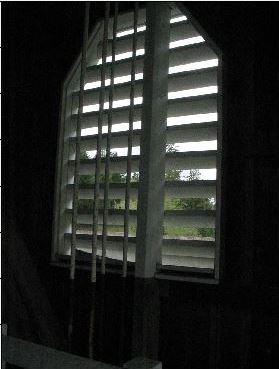} & \includegraphics[height=2cm,keepaspectratio]{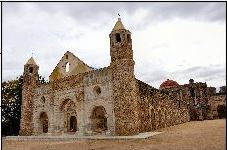} \\
\hline
\rotatebox[origin=c]{90}{Text} & 
\parbox{0.30\linewidth}{\centering Ex Convento at Cuilapan De Guerrero completed in 1555} & 
\parbox{0.30\linewidth}{\centering The Dorena Bridge has these neat windows , which kind makes you feel like you are in a house !} & 
\parbox{0.30\linewidth}{\centering Ex Convento at \textcolor{red}{Dorena Bridge} completed in 1555} \\
\hline
\rotatebox[origin=c]{90}{Location} & 
\parbox{0.30\linewidth}{\centering Mexico, Cuilapam de Guerrero, Oaxaca, Cuilapam de Guerrero. GPS: 16.992657, -96.779133} & 
\parbox{0.30\linewidth}{\centering United States, Lane, Oregon, Row River. GPS: 43.737553, -122.883646} & 
\parbox{0.30\linewidth}{\centering \textcolor{red}{United States, Lane, Oregon, Row River}. GPS: \textcolor{red}{43.737553, -122.883646}} \\
\hline
\hline
\multirow{1}{*}[9ex]{\rotatebox[origin=c]{90}{Image}} & 
\includegraphics[height=2cm,keepaspectratio]{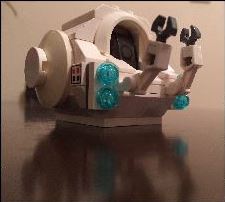} & \includegraphics[height=2cm,keepaspectratio]{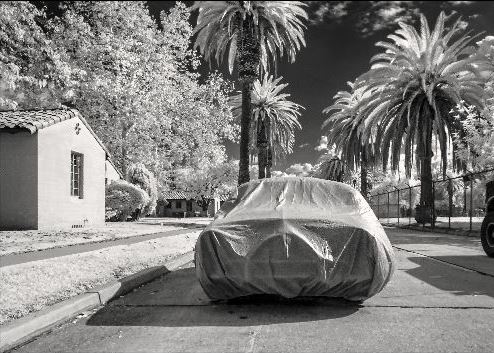} & \includegraphics[height=2cm,keepaspectratio]{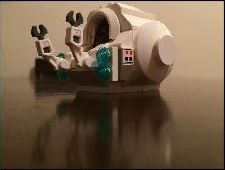} \\
\hline
\rotatebox[origin=c]{90}{Text} & 
\parbox{0.30\linewidth}{\centering The 2001 Space pod modelled in Lego by Dilip} & \parbox{0.30\linewidth}{\centering " It was a dream , and in dreams you have no choices: ..." -- Neil Gaiman , American Gods ( 2001 )} & 
\parbox{0.30\linewidth}{\centering Space pod modelled in Lego by \textcolor{red}{Neil Gaiman}} \\
\hline
\rotatebox[origin=c]{90}{Location} & 
\parbox{0.30\linewidth}{\centering United Kingdom, West Sussex, England, Billingshurst. GPS: 51.024538, -0.450281} & 
\parbox{0.30\linewidth}{\centering United States, Riverside, California, Moreno Valley. GPS: 33.900035, -117.254384} & 
\parbox{0.30\linewidth}{\centering United Kingdom, West Sussex, England, Billingshurst. GPS: 51.024538, -0.450276} \\
\hline
\hline
\multirow{1}{*}[9ex]{\rotatebox[origin=c]{90}{Image}} & 
\includegraphics[height=2cm,keepaspectratio]{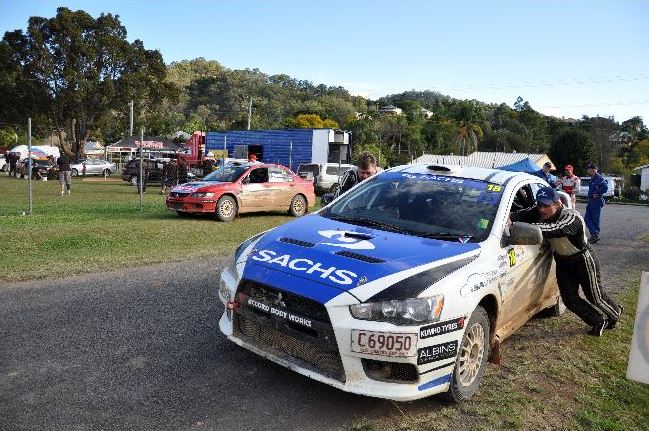} & \includegraphics[height=2cm,keepaspectratio]{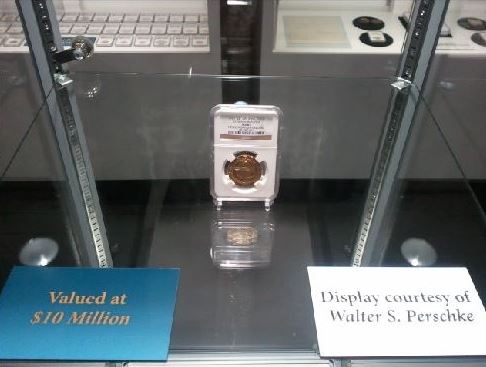} & \includegraphics[height=2cm,keepaspectratio]{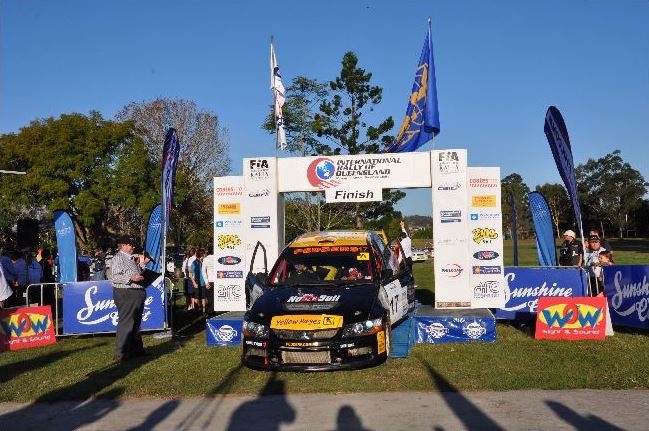} \\
\hline
\rotatebox[origin=c]{90}{Text} & 
\parbox{0.30\linewidth}{\centering The International Rally of Queensland Sunday competition images + podium} & 
\parbox{0.30\linewidth}{\centering \$10 , 000 , 000 00 is the estimated value for the first United States gold coin Numismatic Guaranty Corporation special display ... } & 
\parbox{0.30\linewidth}{\centering The \textcolor{red}{Numismatic Guaranty Corporation} Sunday competition images + podium} \\
\hline
\rotatebox[origin=c]{90}{Location} & 
\parbox{0.30\linewidth}{\centering Australia, Gympie Regional, Queensland, Imbil. GPS: -26.460497, -152.679491} & 
\parbox{0.30\linewidth}{\centering United States, Philadelphia, Pennsylvania, Philadelphia. GPS: 39.953333, -75.156667} & 
\parbox{0.30\linewidth}{\centering Australia, Gympie Regional, Queensland, Imbil. GPS: -26.461131, 152.678117} \\
\hline
\end{tabular}
\caption{Examples of location, person and organization manipulations from the Multimodal Entity Image Repurposing (MEIR) dataset. A manipulation source package provides a foreign entity which is used to replace corresponding entities from the manipulated package. Text in red indicates manipulation. Longer text descriptions have been truncated.}
\label{fig:samples}
\end{center}
\end{table*}

Jaiswal \textit{et al.} \cite{jaiswal_multimedia_2017} presented the Multimodal Information Manipulation (MAIM) dataset of multimedia packages, in which each package contains an image-caption pair. In order to mimic image repurposing, associated captions were swapped with randomly chosen captions---potentially leading to semantically inconsistent repurposing, such as an image of a cat captioned as a car. The MAIM dataset was the first attempt to create a dataset for image repurposing detection research. However, MAIM had several limitations. First, the resulting manipulations are often easy to detect and unlikely to fool a human. Second, the reference dataset provided in MAIM is unlabeled and does not contain entities directly related to manipulated packages. Therefore, there is no explicit use of reference dataset when detecting manipulations. MAIM dataset does not reflect difficult manipulations that are created by carefully altering caption contents to produce semantically consistent packages---for example, leaving the caption unaltered except for replacing the name of an entity. 

In order to address these limitations, we introduce the Multimodal Entity Image Repurposing Dataset (MEIR), which compared to MAIM, contains semantically intelligent manipulations. Multimedia packages in MEIR are manipulated by replacing named entities of corresponding type, instead of entire captions. We consider three named entities: person, organization and location. Unlike MAIM, the reference dataset in MEIR has directly related entities to manipulated packages. It is also labeled to identify this relationship. The process for bringing these improvements in a new dataset is described in the following.

Creating MEIR consisted of three stages---data curation, clustering packages by relatedness, and creating manipulations in a fraction of packages of each cluster. MEIR was created from Flickr images and their associated metadata, collected in the form of packages. To ensure uniformity, packages with missing modalities and non-English text were removed. The remaining packages were preprocessed to remove noise in the form of html tags. Following data collection and preprocessing, packages were clustered by relatedness. Clustering by relatedness helps allocate packages to reference, training, development and test sets, such that related packages are labeled and present in the reference dataset when detecting manipulations. Identifying relatedness before synthesizing manipulations also helps to avoid replacing entities between two related packages. Two packages are considered related when they are geographically located close to each other in terms of real-world location, and the image and text modalities between them are similar. Clustering is done in two stages: clustering by location followed by refining the clusters based on image and text similarities between packages. Initial clusters of packages with neighboring global positioning system (GPS) coordinates were created by measuring the proximity of GPS coordinates up to two decimal places of accuracy. This limited the maximum distance between two related packages to approximately 1.3 kilometers. A relaxed distance constraint between two packages was kept to ensure better recall in the first stage of clustering. In the second stage, clusters were refined based on image and text similarity between packages. Similarity was measured by scoring image and text feature representations of two packages. Feature representations used were VGG19 \cite{simonyan_very_2014}, features pretrained on imagenet and averaged word2vec \cite{mikolov_distributed_2013} embeddings for image and text respectively. Cosine similarity was used to measure similarity. Selection of a suitable threshold on the similarity score was based on a sample of 200 package pairs which was manually annotated for relatedness.

In order to create manipulations, we applied named entity recognition for identifying and replacing entities between packages. We used StanfordNER's \cite{finkel_incorporating_2005} three-class model for identifying person name, location and organization entities in each package. Entities of corresponding type were replaced between two packages from unrelated clusters. All instances of that entity type were manipulated to "cover-up" potential inconsistencies within a manipulated package. As an example, when making a location manipulation of \emph{London} to \emph{Paris}, any mention of \emph{England} should also be manipulated along with the GPS coordinates. Approximately, a quarter of each cluster is manipulated. The reference dataset is allocated to approximately half of the packages from each cluster, such that each package is unmanipulated. The remaining half of each cluster is an equal mix of manipulated and unmanipulated packages and is used to create the train-test-validation split. The reference dataset is common to train, test and validation sets.

The reference dataset contains 82,156 untampered related packages. There are 57,940 packages in the manipulated dataset out of which 28,976 packages are split into 14,397 location, 7,337 person and 7,242 organization manipulations. The manipulated dataset is divided into 40,940, 7,000 and 10,000 packages for training, validation and test, respectively. Each package consists of image, location and text modalities. Location comprises country, county, region and locality names, in addition to GPS coordinates. Text is the user generated description related to the image. The dataset also contains an additional modality---time-stamp which contains the date and time associated with each package. Packages are spread across 106 countries with English-speaking countries contributing the majority.

Examples from MEIR are presented in Table \ref{fig:samples} showing location, person and organization manipulation. It is important to note that a human is unlikely to see through these manipulations. This understanding reinforces the need for a reference dataset containing directly related information.
\section{Image Repurposing Detection}\label{sec:method}

\begin{table}[t]
\small
\centering
\begin{center}
\begin{tabular}{|c|c|c|c|}
\hline
\textbf{Modality} & \textbf{Manipulated} & \textbf{Unmanipulated} & \textbf{Overall} \\
\hline
\hline
GPS coordinates & 47.0\% & \textbf{95.0\%} & 71.0\% \\
Image & 65.9\% & 65.6\% & 65.8\% \\
Text & 60.0\% & 77.2\% & 68.6\% \\
Image+GPS+Text & \textbf{66.1\%} & 93.6\% & \textbf{79.9\%} \\
\hline
\end{tabular}
\end{center}
\caption{Top-1 retrieval accuracy of related packages when a query package is manipulated or unmanipulated. Using all modalities gives the best performance for retrieval.}
\label{table4}
\end{table}

\begin{figure}[t]
\centering
\includegraphics[width=0.90\linewidth,keepaspectratio]{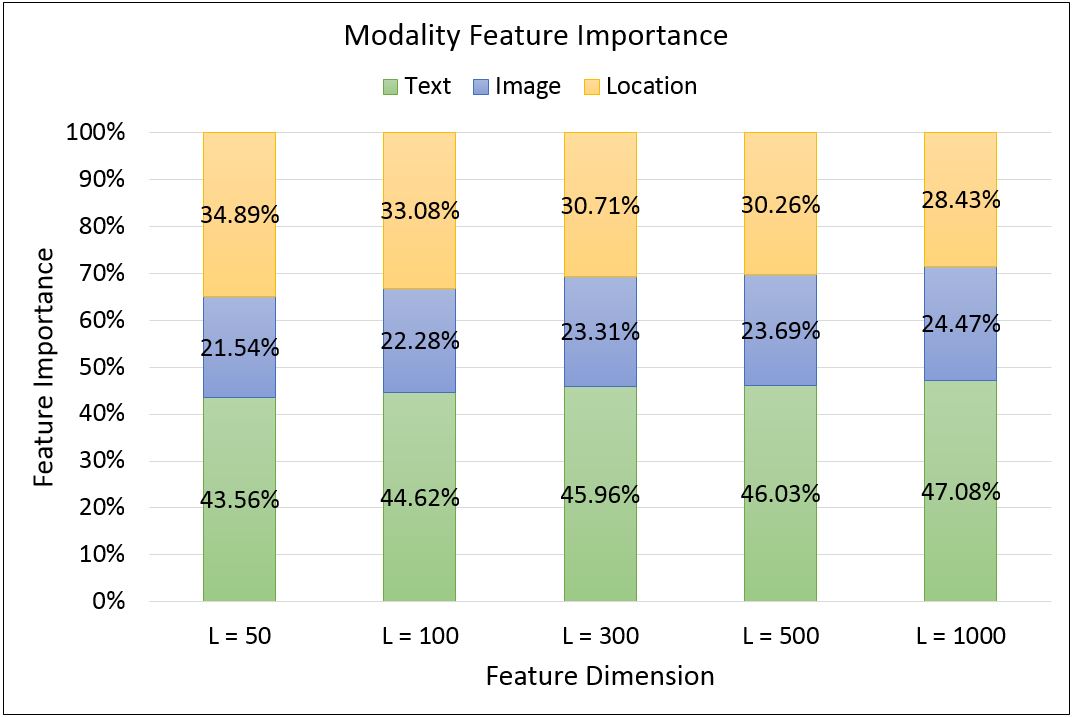}
\caption{Average feature importance of each modality with varying feature dimension. All modalities are found to contribute towards manipulation detection.}
\label{fig:modality}
\end{figure}

\begin{figure*}[t]
\centering
\includegraphics[width=1\textwidth]{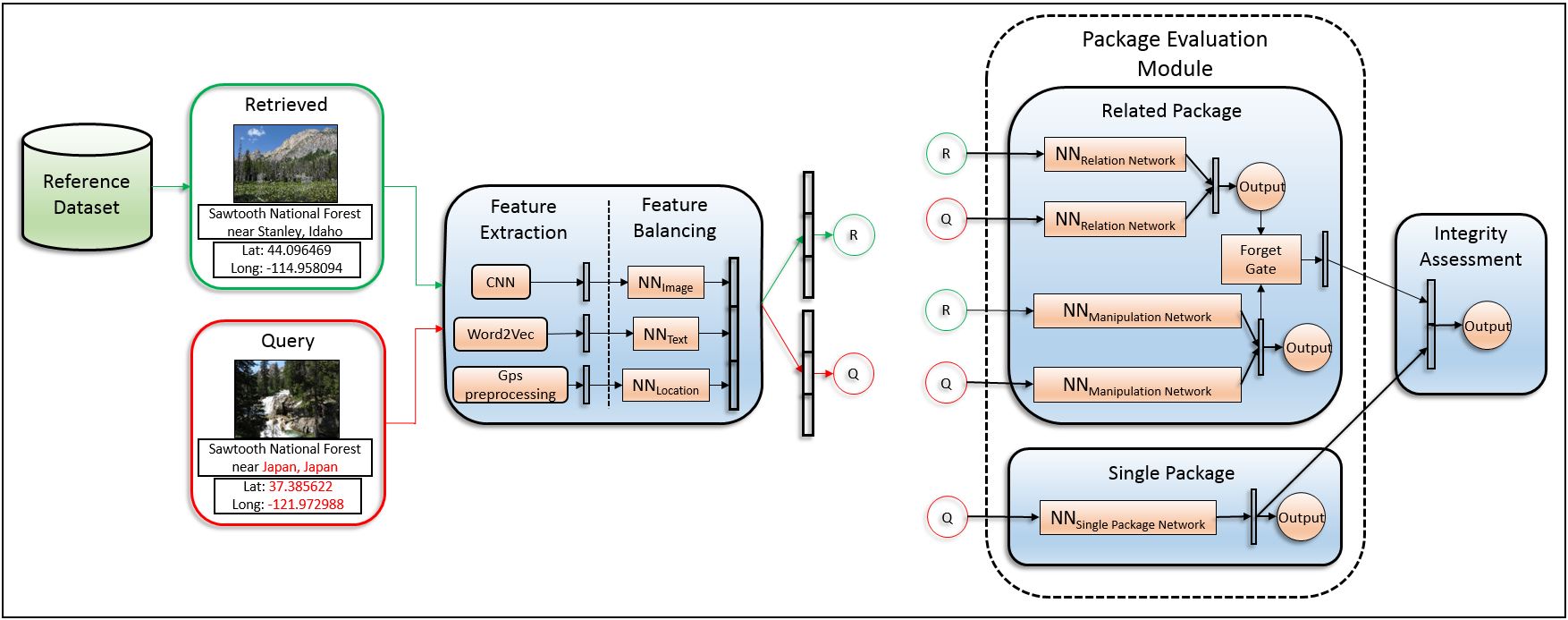}
\caption{Model overview. A potentially related package is retrieved from the reference dataset. Feature extraction and balancing modules create a concatenated feature vector. The package evaluation module consists of related and single package modules. All $NN_{i}$ layers represent a single, dense, fully connected layer with ReLU activation.}
\label{fig:architecture}
\end{figure*}

Manipulation detection systems often retrieve one or more related images or packages from a reference dataset first, followed by comparing the query package to the retrieved one to determine likelihood of manipulation.

Since different modalities in a given package (e.g., image, text, location, etc.) contribute differently to both retrieval and manipulation detection, we start by assessing the importance of each modality to guide the design of the manipulation detection approach. 

\subsection{Modality Importance}\label{sec:retrieval}

Image manipulators are assumed to leave most of the content of the package unchanged, and only change a small number of package modalities (e.g., only location) to make the manipulation subtle and hard to detect. Therefore, the query and retrieved packages are assumed to have largely overlapping information. 

We use similarity scoring to retrieve a package from reference dataset $\mathbb{R}$ using every modality for every query package. Let $f_{qm}$ and $f_{rm}$ be the feature vectors for modality $m$ in query $q$ and reference $r$ packages, respectively, and $s(., .)$ is the similarity metric. The top retrieved package $r^{*}$ is identified as shown in Equation \ref{eq:sim}.
\begin{equation}\label{eq:sim}
\displaystyle r^{*}=\argmax_{r\in \mathbb{R}}  \sum_m s({f_{qm},f_{rm})}
\end{equation}
VGG19 \cite{simonyan_very_2014} pretrained on imagenet \cite{russakovsky_imagenet_2015} and averaged word2vec \cite{mikolov_distributed_2013} features are used for images and text respectively. We use cosine distance to measure package similarities. Experimental results are presented in Table \ref{table4}, which shows the top-1 retrieval results for both manipulated and unmanipulated partitions of the query subset of MEIR. The results illustrate that using all modalities for retrieving top-1 related package provides the best retrieval results. 

Further, in order to estimate the importance of each modality for manipulation detection, we devise a classification experiment followed by feature importance measurement. The model comprises Gaussian random projection for dimensionality reduction followed by random forest \cite{breiman_random_2001} for classification. Random projections have been used previously for dimensionality reduction and in classification problems \cite{bingham_random_2001}\cite{dasgupta_experiments_2000}. Each modality in query and related packages is reduced to a common feature dimension $L$, and features from all modalities are concatenated. A simple random forest classifier is trained on the resulting feature vector for manipulation detection. Feature importance of each dimension is measured using Gini impurity across all trees as described in \cite{breiman_classification_1984} and implemented in \cite{pedregosa_scikit-learn:_2011}. Averaged modality importance is shown in Figure \ref{fig:modality}. Each experiment configuration is repeated for 30 trials and averaged results are presented. As shown in Figure \ref{fig:modality}, all modalities contribute significantly to manipulation detection, at all feature dimensions. 

\subsection{Deep Manipulation Detection Model}

Broadly speaking, there are two approaches for manipulation detection. The first approach depends on assessing the coherency of the content of the query package, without using any reference datasets, e.g., by matching caption against the image or vice versa \cite{jaiswal_multimedia_2017}. The second approach assumes the existence of a relatively large reference dataset, and assesses the integrity of the query package by comparing it to one or more packages retrieved from the reference dataset. The main advantage of the second approach is when the manipulations are semantically consistent. The proposed method in this paper belongs to the second approach, since the information in the query package is potentially manipulated and requires external information for validation.

We propose a deep multimodal, multi-task learning model for image repurposing detection,  as illustrated in Figure \ref{fig:architecture}. The proposed model consists of four modules: (1) feature extraction, (2) feature balancing, (3) package evaluation and (4) integrity assessment. The model takes a query package and the top-1 related package, retrieved from a reference dataset, as discussed in Section \ref{sec:retrieval}.

Both query and reference dataset packages are assumed to contain image, text and GPS coordinates. Images are encoded using VGG19 \cite{simonyan_very_2014} pretrained on imagenet \cite{russakovsky_imagenet_2015}. GPS coordinates are a compact numerical representation of location; therefore we normalize them without any further processing. Finally, text is represented using averaged word2vec \cite{mikolov_distributed_2013} features for all words in a sentence. We also explore performance improvement using an attention model over words instead of averaging word features.

Features from different modalities have widely different dimensionalities (4096D for image, 300D for text and 2D for location). As shown before in Section \ref{sec:retrieval}, varying the dimensionality of each modality does not significantly change its importance. In order to ensure features from all modalities are balanced, all features are transformed to 300D feature vectors (similar to word2vec text features) and concatenated into a single feature vector. Neural layers are used to transform feature dimensions. 

\begin{table}[t]
\small
\centering
\begin{center}
\begin{tabular}{|c|cccccc|}
\hline
\textbf{Component} & \multicolumn{6}{c|}{\textbf{Combination}} \\
\hline
\hline
Single Package & \checkmark & \checkmark & \checkmark & \checkmark & \checkmark & \checkmark\\
Related Package &  & \checkmark & \checkmark & \checkmark & \checkmark & \checkmark \\
Multitask-loss & & & \checkmark & \checkmark & \checkmark & \checkmark \\
Feature Balancing & & & & \checkmark & \checkmark & \checkmark \\
Attention in Text &  &  &  &  & \checkmark & \checkmark \\
Learnable Forget Gate & & & & & & \checkmark \\
\hline
\hline
\textbf{Metric} & \multicolumn{6}{c|}{\textbf{Scores}} \\
\hline
\hline
$F_1$ tampered & 0.58 & 0.59 & 0.61 & 0.76 & 0.80 & \textbf{0.83} \\
$F_1$ clean & 0.60 & 0.60 & 0.65 & 0.78 & 0.82 & \textbf{0.84} \\
AUC & 0.64 & 0.65 & 0.70 & 0.86 & 0.89 & \textbf{0.91} \\
\hline
\end{tabular}
\end{center}
\caption{This table justifies different design choices of the model. A \checkmark indicates the presence of a component. Model architecture has been optimized on the development set.}
\label{table5}
\end{table}

The core of the proposed model is the package evaluation module, which consists of related package and single package sub-modules. As shown in Figure \ref{fig:architecture}, the related package sub-module consists of two Siamese networks. The first network is a \emph{relationship classifier} that verifies whether the query package and top-1 package are indeed related, while the second network is a \emph{manipulation detection} network that determines whether the query package is a manipulated version of the top-1 retrieved package. Since manipulation detection is dependent on the relatedness of the two packages, the relationship classifier network controls a forget gate which scales the feature vector of the manipulation detection network down to zero if the two packages are unrelated. Two designs are considered for a forget gate. The first is a dot product between the output of \emph{relationship classifier} and \emph{manipulation detection feature vector}. This formulation of the forget gate is not learnable and scales all dimensions of the manipulation detection feature vector indiscriminately. An alternative is a learnable forget gate similar to LSTMs \cite{hochreiter_long_1997}. If $x$ is the relationship classifier output, $y_{inp}$ the manipulation detection feature vector, $w$ the forget gate weight matrix and $b$ the bias vector, then the output feature vector $y_{out}$ is given by Equation \ref{eq:gate} where $*$ is the Hadamard product and $\cdot$ is matrix multiplication. For a learnable forget gate, the feature vector of the relationship classifier is used as input. Choice of gate is justified ahead.
\begin{equation}\label{eq:gate}
y_{out} = y_{inp}*\sigma(w\cdot x+b)
\end{equation}
In the meantime, a single package module verifies the coherency (i.e., integrity) of the query package alone.\footnote{Similar to approaches that do not depend on a reference dataset for integrity assessment} The single package module is a feedforward network that takes the balanced feature vector of the query package as input and performs feature fusion to present a 100 dimensional feature vector.

\begin{figure}[t]
\centering
\includegraphics[height=4cm,keepaspectratio]{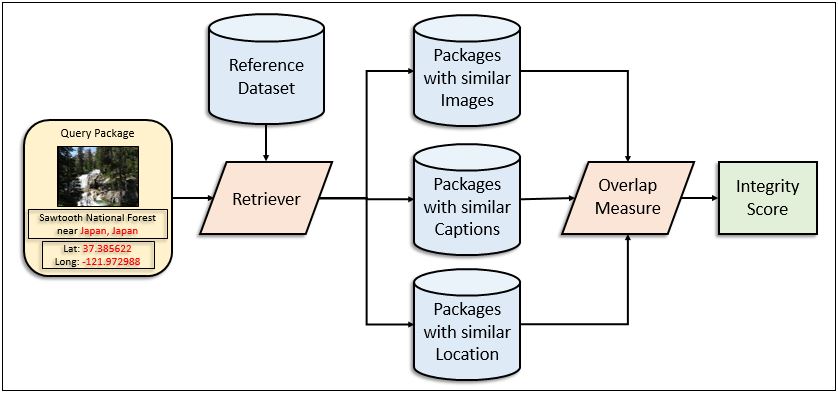}
\caption{Our baseline semantic retrieval system (SRS) retrieves similar concepts via each modality and uses an overlap between retrieved packages as an indicator for integrity score. Jacardian index is used to measure overlap.}
\label{fig:baseline}
\end{figure}

\begin{table}[t]
\small
\begin{center}
\begin{tabular}{|c|c|c|c|c|}
\hline
\multirow{2}{*}{\textbf{Modalities}} & \multirow{2}{*}{\textbf{Method}} & \multicolumn{3}{c|}{\textbf{MEIR}} \\
\cline{3-5}
 & & $F_1$ tampered & $F_1$ clean & AUC  \\
 \hline\hline
Image + & VSM \cite{jaiswal_multimedia_2017} & 0.56 & 0.63 & 0.60 \\
Text & Our Method & \textbf{0.73} & \textbf{0.76} & \textbf{0.81} \\
\hline\hline
Image + & VSM \cite{jaiswal_multimedia_2017} & 0.63 & 0.59 & 0.65 \\
Text + & GCCA & 0.75 & 0.71 & 0.81 \\
GPS & SRS & 0.66 & 0.37 & 0.71 \\
& Our Method & \textbf{0.80} & \textbf{0.80} & \textbf{0.88} \\
\hline
\end{tabular}
\end{center}
\caption{We present results on MEIR. Experiments are performed in two settings---with image and text modalities, and image, location and text modalities. This is done for a fairer comparison with VSM which was originally designed for the image-text dataset.}
\label{table1}
\end{table}

To ensure the overall network behaves as intended, we use multitask learning in the two submodules. Relationship and manipulation classifiers are trained with corresponding labels. To avoid conflict in training, the manipulation classifier in related package submodule has a third label \emph{unknown} apart from \emph{manipulated} and \emph{unmanipulated}, when the retrieved package is unrelated. All neural layers in the package evaluation module are dense, fully connected 100 dimensional layers.

The integrity assessment module concatenates feature vectors from both related and single package modules for manipulation classification. A neural network classifier is trained on the concatenated feature vector with no hidden layers.

The model has multiple components that need to be validated before we evaluate it against other methods and baselines. As discussed previously, there are also choices involved with some components of the model---attention over averaged word2vec features for text feature extraction and learnable over non-learnable forget gate. These choices are tuned on the development set with a set of ablation experiments. The experiments add different components of the model sequentially and notice a performance increase  justifying their addition. By default, averaged word2vec features and non-learnable forget gates are used, unless otherwise mentioned. Experimental results are shown in Table \ref{table5}. A \checkmark indicates the presence of a component. Improvement of 0.16 AUC from the feature balancing layer can be further broken down into 0.04 AUC from increased depth and 0.12 AUC from matching feature dimensions of different modalities. We hypothesize that dimensionality matching prevents modalities with higher dimension (e.g., image) from dominating over low dimension modalities (e.g., GPS). A model comprising all discussed components, with attention over text features and a learnable forget gate, gives the best performance. We use this configuration in all future evaluations.

The model is trained end to end. We use Adam optimizer in all examples with a learning rate of 0.001. All nonlinear transformations use the ReLU activation function. We use Keras with tensorflow backend. Undiscussed parameters are set to default values.

\section{Experimental Evaluation}\label{sec:evaluation}

\begin{table*}[t]
\small
\begin{center}
\begin{tabular}{|c|c|c|c|c|c|c|c|c|c|}
\hline
\multirow{2}{*}{\textbf{Method}} & \multicolumn{3}{c|}{\textbf{MAIM}} & \multicolumn{3}{c|}{\textbf{Flickr30K}} & \multicolumn{3}{c|}{\textbf{MS COCO}} \\
\cline{2-10}
 & $F_1$ tampered & $F_1$ clean & AUC & $F_1$ tampered & $F_1$ clean & AUC & $F_1$ tampered & $F_1$ clean & AUC \\
 \hline\hline
 MAE \cite{jaiswal_multimedia_2017} & 0.49 & 0.49 & - & 0.49 & 0.50 & - & 0.5 & 0. 48 & - \\
BiDNN \cite{jaiswal_multimedia_2017} & 0.52 & 0.52 & - & 0.63 & 0.62 & - & 0.76 & 0.77 & - \\
VSM \cite{jaiswal_multimedia_2017} & 0.75 & 0.77 & - & \textbf{0.89} & \textbf{0.88} & - & \textbf{0.94} & \textbf{0.94} & - \\
Our Method - SPA & \textbf{0.78} & \textbf{0.78} & \textbf{0.87} & 0.88 & \textbf{0.88} & 0.95 & 0.92 & 0.92 & 0.96 \\
\hline
\end{tabular}
\end{center}
\caption{We present results on MultimodAl Information Manipulation (MAIM) dataset. MAIM has subjective image caption pairs. We train our method without Inter-package task module since there are no related packages to leverage in MAIM. We compare against methods presented in \cite{jaiswal_multimedia_2017}.}
\label{table3}
\end{table*}

\begin{table}[t]
\small
\begin{center}
\begin{tabular}{|c|c|c|c|c|}
\hline
\multirow{2}{*}{\textbf{Missing Modality - Training}} & \multicolumn{4}{c|}{\textbf{Missing Modality - Test}} \\
\cline{2-5}
& None & Image & Text & Location \\
\hline
None & \textbf{0.88} & 0.77 & 0.66 & 0.63 \\
Image & 0.87 & 0.85 & - & - \\
Text & 0.87 & - & 0.75 & - \\
Location & 0.85 & - & - & \textbf{0.82} \\
All & \textbf{0.88} & \textbf{0.86} & \textbf{0.76} & 0.79 \\
\hline
\end{tabular}
\end{center}
\caption{Scores with missing modalities in retrieved packages. \textit{None} and \textit{All} refer to no missing and all missing modalities respectively. The training scheme improves performance for all missing modality scenarios.}
\label{table2}
\end{table}

\begin{table*}[t]
\small
\makegapedcells
\begin{center}
\begin{tabular}{|c|c|c|c|}
\cline{2-4}
\multicolumn{1}{c|}{} & \multicolumn{2}{c|}{\textbf{Reference Dataset}} & \textbf{Manipulated Dataset} \\
\cline{2-4}
\multicolumn{1}{c|}{} & \textbf{Related Package} & \textbf{Manipulation Source Package} & \textbf{Repurposed Query Package} \\
\cline{2-4}
\hline
\multirow{1}{*}[9ex]{\rotatebox[origin=c]{90}{Image}} &
\includegraphics[height=2cm,keepaspectratio]{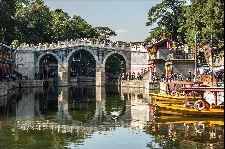} & \includegraphics[height=2cm,keepaspectratio]{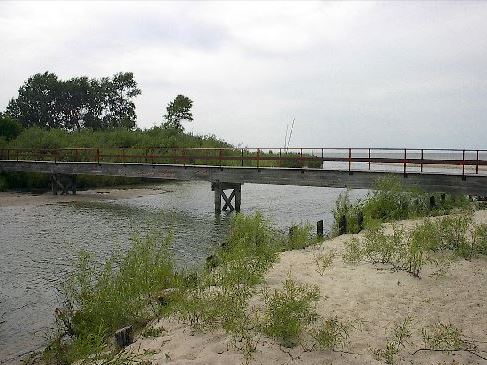} & \includegraphics[height=2cm,keepaspectratio]{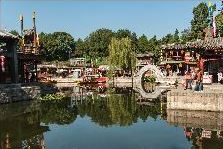} \\
\hline
\rotatebox[origin=c]{90}{Text} &
\parbox{0.30\linewidth}{\centering Sommerpalast Suzhou Street} & \parbox{0.30\linewidth}{\centering Grand Beach , Lake Winnipeg , Manitoba , Canada} & 
\parbox{0.30\linewidth}{\centering \textcolor{red}{Canada}} \\
\hline
\rotatebox[origin=c]{90}{Location} &
\parbox{0.30\linewidth}{\centering China, Beijing, Beijing, Beijing. GPS: 40.000826, 116.268825} & 
\parbox{0.30\linewidth}{\centering Canada, Manitoba, Manitoba, Grand Beach. GPS: 50.562316, --96.614059} & 
\parbox{0.30\linewidth}{\centering \textcolor{red}{Canada, Manitoba, Manitoba, Grand Beach}. GPS: \textcolor{red}{50.562316, --96.614059}} \\
\hline
\hline
\multirow{1}{*}[9ex]{\rotatebox[origin=c]{90}{Image}} &
\includegraphics[height=2cm,keepaspectratio]{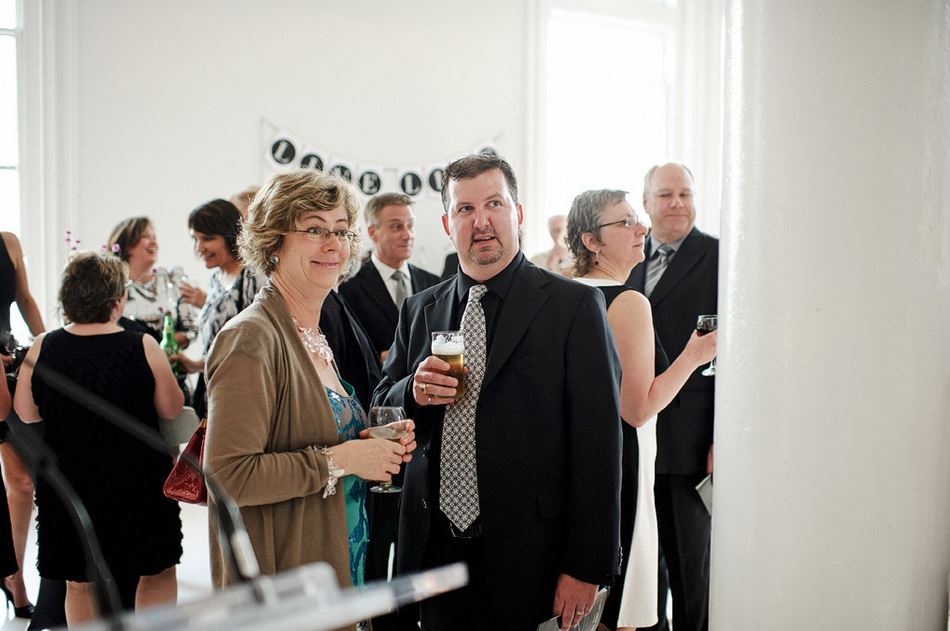} & \includegraphics[height=2cm,keepaspectratio]{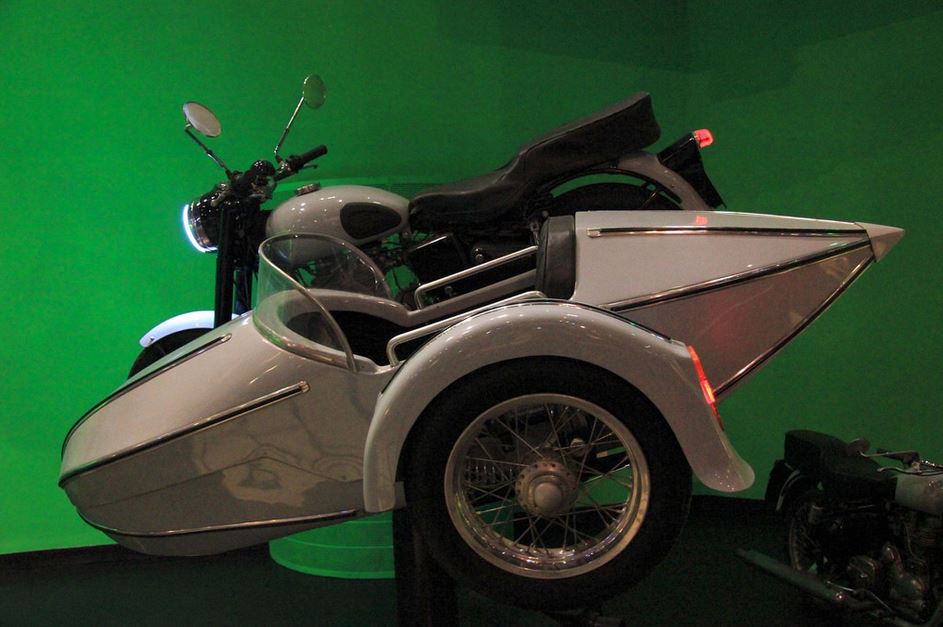} & \includegraphics[height=2cm,keepaspectratio]{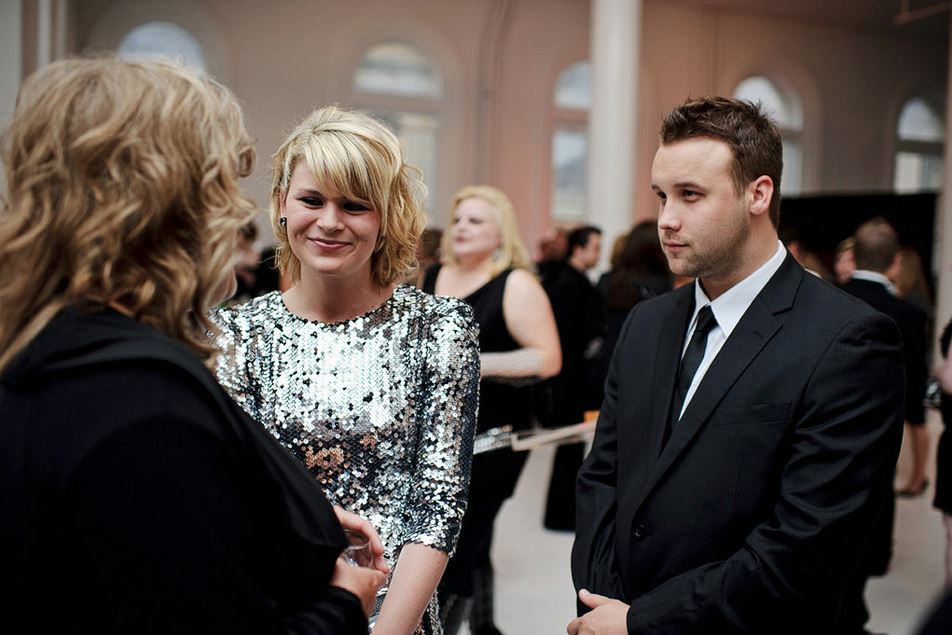} \\
\hline
\rotatebox[origin=c]{90}{Text} &
\parbox{0.30\linewidth}{\centering Live Life Awards Image by Sean McGrath Photography} & 
\parbox{0.30\linewidth}{\centering Film Effects - The Vault CartLike many effects seen in the films ... Ollivanders wand shop , Flourish and Blotts , the ... } & \parbox{0.30\linewidth}{\centering Live Life Awards Image by \textcolor{red}{Blotts} Photography } \\
\hline
\rotatebox[origin=c]{90}{Location} &
\parbox{0.30\linewidth}{\centering Canada, Saint John, New Brunswick, Saint John. GPS: 45.271428, -66.061981} & 
\parbox{0.30\linewidth}{\centering United Kingdom, Hertfordshire, England, Watford. GPS: 51.693761, -0.422329} & 
\parbox{0.30\linewidth}{\centering Canada, Saint John, New Brunswick, Saint John. GPS: 45.271428, -66.061981} \\
\hline
\hline
\multirow{1}{*}[9ex]{\rotatebox[origin=c]{90}{Image}} &
\includegraphics[height=2cm,keepaspectratio]{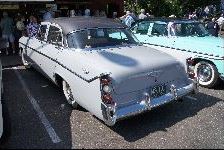} & \includegraphics[height=2cm,keepaspectratio]{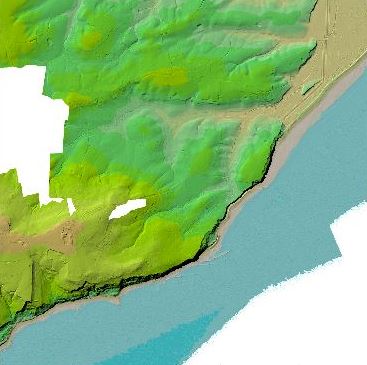} & \includegraphics[height=2cm,keepaspectratio]{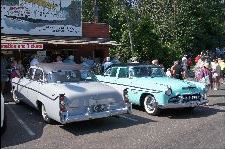} \\
\hline
\rotatebox[origin=c]{90}{Text} &
\parbox{0.30\linewidth}{\centering 1st Combined Convention 28th Annual National DeSoto Club Convention \& 44th Annual Walter P Chrysler Club ConventionJuly 17 - 21 , 2013Lake Elmo , MinnesotaClick link below for more car } & \parbox{0.30\linewidth}{\centering Quick-Look Hill-shaded Colour Relief Image of 2014 2m LIDAR Composite Digital Terrain Model ( DTM ) Data supplied by Environment Agency under the ...} & 
\parbox{0.30\linewidth}{\centering 1st Combined Convention 28th \textcolor{red}{Environment Agency} ConventionJuly 17 - 21 , 2013Lake Elmo , MinnesotaClick link below for more car pictures:} \\
\hline
\rotatebox[origin=c]{90}{Location} &
\parbox{0.30\linewidth}{\centering United States, Chisago, Minnesota, Taylor Falls, GPS: 45.401366, -92.651065} & 
\parbox{0.30\linewidth}{\centering United Kingdom, East Sussex, England, Fairlight Cove, GPS: 50.881552, 0.664108} & 
\parbox{0.30\linewidth}{\centering United States, Chisago, Minnesota, Taylor Falls, GPS: 45.401366, -92.651065} \\
\hline
\end{tabular}
\caption{Location, person and organization manipulation examples in order. Our model identifies each of these examples correctly from test set. In location manipulation China is replaced by Canada, while unrelated names are swapped for person manipulation. The Annual National Club Convention \& 44th Annual Walter P Chrysler Club has been replaced by Environment Agency for organization manipulation. Longer text descriptions have been truncated in examples.}
\label{fig:examples}
\end{center}
\end{table*}

Detection of image repurposing is a relatively new research area without established methods for evaluation. In this paper we propose using the area under receiver operating characteristic curve (AUC) and  also  $F_{1}$ scores for evaluating performance.

We compare the performance of our model against the state-of-the-art in \cite{jaiswal_multimedia_2017}. The visual semantic model (VSM) is the best encoding model in the anomaly detection framework presented in \cite{jaiswal_multimedia_2017}. Further, we also compare our performance against two baselines: the first is our baseline semantic retrieval system (SRS) shown in Figure \ref{fig:baseline}. SRS retrieves similar packages corresponding to each modality using cosine distance. The output integrity score is the measured overlap between retrieved packages. Intuitively, this overlap indirectly measures whether modalities in a query package point to the same related packages in the reference dataset. Since each modality will retrieve similar packages, a rogue modality will retrieve packages pertaining to a different event from the reference dataset. If modalities in the query package are consistent with information in the reference dataset, the overlap between retrieved packages will be significant. We use the Jaccard index for measuring overlap between retrieved packages. The second baseline approach uses generalized canonical correlation analysis (GCCA) \cite{kettenring_canonical_1971} for feature embedding and random forest for classification. GCCA transforms multimodal features with varying feature dimensions into a common embedding space and has been used for classification with multimodal features \cite{shen_generalized_2014}\cite{sun_generalized_2013}.

We also explore the performance of our model without retrieving the top-1 package from a reference dataset. In this setting we remove the related package module. We call this version of our model the single package assessment (SPA) in all figures. This variation enables us to analyze the importance of related package retrieval and compare the performance on MEIR and MAIM datasets.

The results of our experiments on MEIR are presented in Table \ref{table1}, which illustrates that  our model performs better than the baselines and VSM  \cite{jaiswal_multimedia_2017}. It should be noted that VSM was originally designed for image and text modalities. We extend it with location modality. We also evaluate our model with image and text modality for a complete comparison with VSM. Under both circumstances our model performs better. The extended version of VSM with location modality performs better than the original version. Table \ref{fig:examples} shows examples of location, person and organization manipulation from the test set which our model identifies correctly.

\citet{jaiswal_multimedia_2017} evaluate VSM and other methods on three datasets---MAIM, MSCOCO and Flickr30K. MAIM has subjective image caption pairs and is more challenging compared to Flickr30K and MSCOCO, which have objective image caption pairs. However, these datasets do not have related content in the reference dataset. This does not help the related package module in our dataset. We therefore compare the single package assessment (SPA) version of our model without retrieving any query package. They use a representation learning and outlier detection method for comparison. We compare against all encoding methods shown in the referenced paper. They do not provide AUC scores for their method. In Table \ref{table3}, SPA gives superior performance on MAIM and competitive performance on Flickr30K and MS COCO.

A reference dataset may not have all modalities present, which makes missing modalities from retrieved packages an additional problem of interest. Without modifying model architecture, the problem can be tackled by improving the training scheme. During training a modality is deliberately removed from some retrieved samples and represented as a zero vector. Table \ref{table2} shows results of training with a 20\% missing modality. At test time, the corresponding modality is completely removed from all samples. The new training scheme improves performance on missing modalities when compared to simple model training, while maintaining competitive performance on samples with no missing modalities.

\section{Conclusion}

With today's increase in fake news, image repurposing detection has become an important problem. It is also a relatively unexplored research area. The scope of image repurposing was expanded to semantically consistent manipulations which are more likely to fool people. We presented the MEIR dataset with intelligent and semantically consistent manipulations of location, person and organization entities. Our end-to-end deep multimodal learning model gives state-of-the art performance on MEIR and MAIM. The model is also shown to be robust to missing modalities with a proper training scheme.

\begin{acks}

This work is based on research sponsored by the Defense Advanced Research Projects Agency under agreement number FA8750-16-2-0204. The U.S. Government is authorized to reproduce and distribute reprints for governmental purposes  notwithstanding any copyright notation thereon. The views and conclusions contained herein are those of the authors and should not be interpreted as necessarily representing the official policies or endorsements, either expressed or implied, of the Defense Advanced Research Projects Agency or the U.S. Government.

\end{acks}
\clearpage

\bibliographystyle{ACM-Reference-Format}
\balance
\bibliography{ImageRepurposing}

\end{document}